\documentstyle[aps,prb,amsmath,amssymb,psfig]{revtex}
\begin{document}
\wideabs{
\title{Quantum effects on the BKT phase
transition of two-dimensional Josephson arrays}
\author{Alessandro Cuccoli, Andrea Fubini,
        Valerio Tognetti}
\address{Dipartimento di Fisica dell'Universit\`a di Firenze
    and Istituto Nazionale di Fisica della Materia (INFM),
    \\ Largo E. Fermi~2, I-50125 Firenze, Italy}
\author{Ruggero Vaia}
\address{Istituto di Elettronica Quantistica
    del Consiglio Nazionale delle Ricerche,
    via Panciatichi~56/30, I-50127 Firenze, Italy,
    \\ and Istituto Nazionale di Fisica della Materia (INFM)}
\date{\today}
\maketitle

\begin{abstract}
The phase diagram of two dimensional Josephson arrays is studied by means
of the mapping to the quantum XY model. The quantum effects onto the
thermodynamics of the system can be evaluated with quantitative accuracy
by a semiclassical method, the {\em pure-quantum self-consistent harmonic
approximation}, and those of dissipation can be included in the same
framework by the Caldeira-Leggett model. Within this scheme, the critical
temperature of the superconductor-to-insulator transition, which is a
Berezinskii-Kosterlitz-Thouless one, can be calculated in an extremely
easy way as a function of the quantum coupling and of the dissipation
mechanism. Previous quantum Monte Carlo results for the same model appear
to be rather inaccurate, while the comparison with experimental data leads
to conclude that the commonly assumed model is not suitable to describe in
detail the real system.
\end{abstract}
}

\noindent Two-dimensional {\em Josephson junction arrays}
(JJA)~\cite{jja96,BL96,RJ96} have been recently described by means of the
two-dimensional quantum XY model. These arrays are made by superconducting
islands, where the charge carriers (Cooper pairs) interact with each other
through the Coulomb interaction and move between nearest-neighbor islands
through the Josephson tunneling mechanism. The following action turns out
to describe the system~\cite{AES,Metal}
\begin{eqnarray}
 S[\boldsymbol\phi] &=&\int_0^{\hbar\beta}\!\! du \bigg\{\frac{\hbar^2}2
 \sum_{\boldsymbol{ij}} \frac{C_{\boldsymbol{ij}}}{q^2}~{\dot
 \phi_{\boldsymbol{i}}(u)\,\dot \phi_{\boldsymbol{j}}(u)}
\notag \\
 & &~~~~~~~ + J\sum_{\langle{\boldsymbol{ij}}\rangle}
 \Big[1-\cos\big((\phi_{\boldsymbol{i}}(u)
 -\phi_{\boldsymbol{j}}(u)\big)\Big] \bigg\}~;
\label{e.SXY}
\end{eqnarray}
it is a variant of the well-known quantum XY model and therefore a
Berezinskii-Kosterlitz-Thouless (BKT) transition is expected to occur
at some finite temperature: its value is affected by quantum effects,
whose importance depends on the relative weight of the two terms in
the action.  The first one represents the Coulomb interaction between
Cooper pairs with charge $q=2e$ and with the capacitance matrix
\begin{equation}
 C_{\boldsymbol{ij}}=C_0\,\Big[\,\delta_{\boldsymbol{ij}}
 +\eta\,\big( z\,\delta_{\boldsymbol{ij}}-
 {\sum}_{\boldsymbol{d}}\delta_{\boldsymbol{i},\boldsymbol{j+d}}
 \big)\Big]~,
\label{e.Cij}
\end{equation}
where $C_0$ and $C_1\equiv{\eta\,C_0}$ are the self- and mutual
capacitances of the islands, and $\boldsymbol{d}$ runs over the vector
displacements of the $z$ nearest-neighbors. The second term describes the
Josephson interaction with coupling $J$ between nearest-neighbor islands
$\langle{\boldsymbol{ij}}\rangle$, being
$\phi_{\boldsymbol{i}}-\phi_{\boldsymbol{j}}$ the phase difference between
the $\boldsymbol{i}$th and the $\boldsymbol{j}$th superconducting island.
>From a quantum mechanical point of view the superconducting phase
operators $\hat{\phi_{\boldsymbol i}}$ are canonically conjugated to the
Cooper-pair number operators $\hat{n}_{\boldsymbol{i}}$,
$[\hat\phi_{\boldsymbol{i}},\hat{n}_{\boldsymbol{j}}]
=i\,\delta_{\boldsymbol{ij}}$.~\cite{Shon90}

The phase diagram of square and triangular lattices of Josephson Junctions
was experimentally~\cite{ZEGM96} investigated and compared with the
results of quantum Monte Carlo (QMC) simulations~\cite{RJ96} of the above
model.

In this paper we present an analytical study of the phase diagram of JJA
based on the effective potential approach, or {\em pure-quantum
self-consistent harmonic approximation}~\cite{CGTVV95+92} (PQSCHA), where
the effect of quantum dissipation is also considered. In mesoscopic
systems, like JJA, environmental effects can modify the physical
properties of the isolated system. In order to study the open system, an
additional term to the action~(\ref{e.SXY}) is thus
inserted~\cite{CL83,Weiss99}:
\begin{equation}
 S_{\rm D}[\boldsymbol\phi] =\frac12 \int_0^{\hbar\beta} \!\!\!\!du
 \int_0^{\hbar\beta} \!\!\!\!du^{\prime} \sum_{\boldsymbol{ij}}
 K_{\boldsymbol{ij}}(u{-}u^{\prime})~\phi_{\boldsymbol i}(u)\,
 \phi_{\boldsymbol j}(u^{\prime})~,
\label{e.SD}
\end{equation}
where the kernel matrix ${\boldsymbol{K}}(u)=\{K_{\boldsymbol{ij}}(u)\}$
is a real symmetric matrix and, as a function of $u$, is even and
periodic,
${\boldsymbol{K}}(u)={\boldsymbol{K}}(-u)={\boldsymbol{K}}(\beta{-}u)$,
and satisfies $\int_0^\beta{du}\,{\boldsymbol{K}}(u)=0$. It contains the
whole information about the environmental coupling, in particular it is
related to the classical damping memory functions
$\gamma_{\boldsymbol{ij}}(t)$ by
\begin{equation}
K_{n,{\boldsymbol{ij}}}=|\nu_n| {\hat\gamma}_{\boldsymbol{ij}}(|\nu_n|)~,
\label{e.Knij}
\end{equation}
where $\nu_n=2\pi{n}/\hbar\beta$ are the Matsubara frequencies,
$\hat\gamma_{\boldsymbol{ij}}(s)$ means the Laplace transform of
$\gamma_{\boldsymbol{ij}}(t)$, and
\begin{equation}
 {\boldsymbol{K}}_n= \int_0^{\hbar\beta}
 du~{\boldsymbol{K}}(u)~\cos\nu_nu
\end{equation}
is the $n$th Matsubara component of the kernel matrix.

In Eq.~(\ref{e.SXY}), the Coulomb term is like a kinetic energy
contribution (with $\boldsymbol{C}/q^2$ as the mass matrix), while the
Josephson term plays the role of the potential energy. For large
capacitive coupling the classical limit is approached and the system
behaves as a classical XY model, displaying a BKT phase
transition~\cite{BKT} at the classical critical temperature. In the
opposite limit, the energy cost for transferring charges between
neighboring islands is high, so the charges tend to be localized and phase
ordering tends to be suppressed at lower temperatures. An important role
in this system is played by the Coulomb interaction range, that can be
quantified by means of the parameter $\eta$, i.e. the ratio between the
mutual- and the self-capacitance of the junctions. The connection between
$\eta$ and the charging interaction range can be found observing that the
latter is proportional to the inverse of the capacitance
matrix~\cite{Metal}. The Josephson coupling of an island with its
neighbors, $Jz$, defines the overall energy scale for the system, while
the characteristic quantum energy scale $\hbar\Omega$ can be identified
considering the bare dispersion relation,
\begin{equation}
 \hbar^2\Omega_{\boldsymbol{k}}^2= \frac{q^2J}{C_0}~
 \frac{z\,\mu_{\boldsymbol{k}}}{1+z\,\eta\,\mu_{\boldsymbol{k}}}~,
\end{equation}
where $\mu_{\boldsymbol{k}}=1{-}\frac1z \sum_{\boldsymbol{d}}
 \cos(\boldsymbol{k{\cdot}d})$, and choosing $\Omega$ as the maximum frequency
\begin{equation}
 \Omega^2 \equiv \max \left\{\Omega_{\boldsymbol{k}}^2\right\}=
 \frac{q^2J}{\hbar^2 C_0} \cdot
 \begin{cases}
  \displaystyle{ \frac{8}{1+8\,\eta} } & \square~ \text{lattice} ~,
 \\
  \displaystyle{ \frac{9}{1+9\,\eta} } & \triangle~ \text{lattice}~.
 \end{cases}
 \label{e.Omega}
\end{equation}
One can define a meaningful quantum coupling parameter, which rules the
importance of `quanticity' in the system, as the ratio between the two
energy scales
\begin{equation}
 g=\frac{\hbar\Omega}{z\,J}~.
 \label{e.g}
\end{equation}

The thermodynamic properties of the system are studied here by means of
the PQSCHA, which was recently extended to treat quantum open
systems~\cite{CFTV99+97}. By means of this approximation scheme the
thermodynamics of the quantum system~(\ref{e.SXY}) with
dissipation~(\ref{e.SD}) can be reduced to an effective classical problem,
where the pure-quantum part of the fluctuations is taken into account at
the self-consistent harmonic level through a temperature-dependent
renormalization coefficient, while nonlinear thermal fluctuations and
quantum harmonic excitations are fully accounted for. Following the
prescription of Ref.~\onlinecite{CFTV99+97} the PQSCHA effective potential
for our system reads
\begin{equation}
 V_{\rm eff}= J_{\rm eff} \sum_{\langle{\boldsymbol{ij}}\rangle}
 \big[1-\cos(\phi_{\boldsymbol{i}}-\phi_{\boldsymbol{j}})\,\big]~,
\label{e.Veff}
\end{equation}
where we do not consider some additive uniform terms, since they do
influence neither thermal averages nor the critical behavior. The
parameter $J_{\rm eff}$ includes the pure-quantum corrections; it
reads
\begin{equation}
J_{\rm eff}(t,g,\eta,\hat\gamma_{\boldsymbol{ij}}) = J\,e^{-\frac12
{\cal D}(t,g,\eta,\hat\gamma_{\boldsymbol{ij}})}~,
\label{e.Jeff}
\end{equation}
where the renormalization coefficient
${\cal{D}}(t,g,\eta,\hat\gamma_{\boldsymbol{ij}})$ represents the
pure-quantum fluctuations of the relative phase between islands,
$\phi_{\boldsymbol{i}}-\phi_{\boldsymbol{i+d}}$, and is a function of the
{\em reduced temperature} $t=(\beta J)^{-1}$, of the quantum coupling
parameter $g$, of the Coulomb interaction range $\eta$, and of the damping
through $\hat\gamma_{\boldsymbol{ij}}$. In the case of a square lattice
(the generalization to other kind of lattices is straightforward) the
explicit expression of the renormalization coefficient is
\begin{equation}
 {\cal D}=  \frac{(1+8\,\eta)t}{2N} \sum_{\boldsymbol{k}}
 \frac{\mu_{\boldsymbol{k}}}{1+4\,\eta\mu_{\boldsymbol{k}}}
 \sum_{n=1}^{\infty}
 \frac{1}{\widetilde{\nu}_n^2+\widetilde{\omega}_{\boldsymbol{k}}^2
 +\widetilde{\kappa}_{n,{\boldsymbol{k}}}}~.
 \label{e.D1}
\end{equation}
where the tilde means that the square frequencies $\nu_n^2$,
$\omega_{\boldsymbol{k}}^2$ and $\kappa_{n,{\boldsymbol{k}}}$ are
expressed in units of $\Omega$, the characteristic frequency scale. They read
\begin{eqnarray}
 & & \widetilde{\nu}_n =\frac{2\pi n\,t}{g}~,~~~~~~~
 \widetilde{\omega}_{\boldsymbol{k}}= \widetilde{\Omega}_{\boldsymbol{k}}
 \,e^{-{\cal D}/4}~,
\nonumber\\
 & &~~~~~~~
 \widetilde{\kappa}_{n,{\boldsymbol{k}}}=
 \frac{g}{8}\,\frac{1+8\,\eta}{1+4\,\eta\mu_{\boldsymbol{k}}}
 \widetilde{K}_{n,{\boldsymbol{k}}}~,
 \label{e.adim}
\end{eqnarray}
where $C_{\boldsymbol{k}}$ and $K_{n,{\boldsymbol{k}}}$ are the Fourier
transforms of~(\ref{e.Cij}) and~(\ref{e.Knij}) respectively. The
renormalization coefficient ${\cal D}(t)$ is very sensitive to the range
of the charging interaction. Indeed, for a fixed value of the quantum
coupling the pure-quantum fluctuations of the phase represented by
${\cal{D}}(t)$ are strongly enhanced when $\eta$ increases, and they
saturate when $\eta\gg{1/\pi{z}}$. This behavior can be explained by
writing the dispersion relation of the linear excitations,
\begin{equation}
 \omega_{\boldsymbol{k}}^2= \frac{q^2J_{\rm eff}}{C_0}
 \,\frac{z\,\mu_{\boldsymbol{k}}}{1+z\,\eta\,\mu_{\boldsymbol{k}}}
 ~~\xrightarrow[\eta \to \infty]
 ~~~\omega_{{}_{\rm E}}^2=\frac{q^2J_{\rm eff}}{C_1}~;
 \label{e.omegak}
\end{equation}
as $\eta$ increases, a larger and larger region of the dispersion relation
tends to the constant frequency $\omega_{{}_{\rm{E}}}$, and the relative
portion of the Brillouin zone where $\omega_{\boldsymbol{k}}$ differ
significantly from $\omega_{{}_{\rm E}}$ shrinks as $(\pi{z}\eta)^{-1}$.
On the other hand, the low-frequency part of the spectrum does not
contribute significantly to the pure-quantum coefficient ${\cal D}(t)$ due
to the absence of the $n=0$ Matsubara term in the summation of
Eq.~(\ref{e.D1}).

This behavior of ${\cal{D}}(t)$ also differs from what was found in
Ref.~\onlinecite{KC95} (where ${\cal{D}}$ is denoted as $-2\ln{g_0}$):
this results from the different choice of the quantum coupling parameter.
Our choice, Eq.~(\ref{e.g}), takes into account the contributions of the
whole capacitance interaction in order to determine a meaningful quantum
energy scale~(\ref{e.Omega}). In particular, while the quantum coupling
parameter chosen in Ref.~\onlinecite{KC95} contains the self-capacitance
term only, our $g$ varies continuously with $\eta$ and, in the limits of
small and large $\eta$, it smoothly connects the two quantum coupling
parameters of Ref.~\onlinecite{RJ96}. We notice that in
Ref.~\onlinecite{KC95} the effective Josephson coupling is erroneously set
to
$J_{\rm{eff}}/J=(1+{\cal{D}}/2)\,e^{-{\cal{D}}/2}\sim{1}-{\cal{D}}^2/8$,
while the correct low-coupling approximation of $V_{\rm{eff}}$ gives
Eq.~(\ref{e.Jeff}), i.e. $J_{\rm{eff}}/J\sim{1}-{\cal{D}}/2$. Therefore in
Ref.~\onlinecite{KC95} the quantum effects turn out to be significantly
underestimated. The same mistake is made in Refs.~\onlinecite{KC90}.

Starting from Eqs.~(\ref{e.Veff}-\ref{e.D1}), the quantum thermodynamic
average of any observable
$\hat{\cal{O}}(\boldsymbol{\hat{n},\hat\phi})$, e.g. a correlation
function or the helicity modulus, can be expressed in terms of its
Weyl symbol $\cal{O}(\boldsymbol{n,\phi})$ by a classical-like formula
given by~\cite{CFTV99+97},
\begin{equation}
 \big\langle \hat{\cal{O}}(\boldsymbol{\hat{n},\hat\phi})\big\rangle
 = \Big\langle\,\big\langle\!\!\big\langle
 {\cal{O}}(\boldsymbol{n,\phi+\xi})
 \big\rangle\!\!\big\rangle \,\Big\rangle_{\rm eff}~,
\end{equation}
where
\begin{equation}
 \langle\,\cdot\,\rangle_{\rm eff} = {\cal{Z}}_{\rm C}^{-1}
 \int d\boldsymbol\phi \, (\,\cdot\,) \,
 e^{-\beta V_{\rm eff}(\boldsymbol\phi)}
\end{equation}
(${\cal{Z}}_{\rm C}=\int d\boldsymbol\phi \, e^{-\beta V_{\rm
eff}(\boldsymbol\phi)}$) and
$\big\langle\!\!\big\langle {\cal{O}}(\boldsymbol{n,\phi+\xi})
\big\rangle\!\!\big\rangle$ denotes the average over the two
independent Gaussian distributions of $\boldsymbol{n}$
and $\boldsymbol{\xi}$: the former accounts for the total fluctuations of the
number variables,  while the latter smears the function
$\cal{O}(\boldsymbol{n,\phi})$
on the scale of the pure-quantum fluctuations $\boldsymbol\xi$ of the
phases.

Within the PQSCHA approach, the BKT critical temperature
$t_{{}_{\rm{BKT}}}$ of the system, can be easily evaluated by the
following self-consistent relation~\cite{CTVV95qxxz}:
\begin{equation}
t_{{}_{\rm{BKT}}}(g,\eta,\hat\gamma_{\boldsymbol{ij}})
=t^{\rm (cl)}_{{}_{\rm{BKT}}}\,
e^{-\frac12 {\cal D}(t_{{}_{\rm{BKT}}},g,\eta,\hat\gamma_{\boldsymbol{ij}})}~,
\end{equation}
where $t^{\rm (cl)}_{{}_{\rm{BKT}}}$ is the BKT critical temperature of
the classical XY model, $t^{\rm (cl)}_{{}_{\rm{BKT}}} =
0.895$~\cite{GB92} for the square lattice and $t^{\rm (cl)}_{{}_{\rm{BKT}}} =
1.36$~\cite{BC94} for the triangular one.

\begin{figure}[t]
\centerline{
\psfig{bbllx=35mm,bblly=32mm,bburx=192mm,bbury=234mm,%
 figure=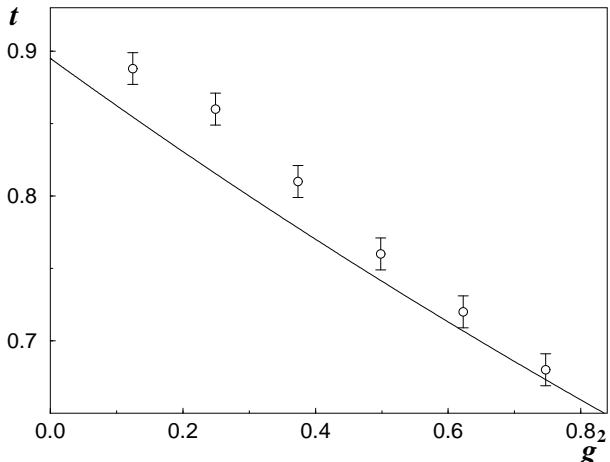,width=80mm,angle=270}}
 \caption{ Phase diagram for the quantum XY model, Eq.~(\ref{e.SXY}), on a
square lattice in the $g^2\!\!-\!t $ plane at fixed value of $\eta=100$.
The line is our result for the undamped system, i.e. $
\Gamma_0=\Gamma_1=0$. The circles are the QMC results of
Ref.~\protect\onlinecite{RJ96}.} \label{f.MC}
\end{figure}

Turning to our results, let us consider firstly the behavior of the
undamped system ($\widetilde{\kappa}_{n,{\boldsymbol{k}}}=0$). Our phase
diagram ($t_{{}_{\rm{BKT}}}$ vs. the square coupling $g^2$) for the square
lattice is compared with quantum Monte Carlo simulations~\cite{RJ96} in
Fig.~\ref{f.MC}. We observe that our approach starts, by construction,
from the correct classical value $t^{\rm (cl)}_{{}_{\rm{BKT}}}$ and
remains valid as long as ${\cal D}\ll 1$: this means that the reduction of
the transition temperature due to the quantum effects can be considered
reliable if less than about $30\%$ of $t^{\rm (cl)}_{{}_{\rm{BKT}}}$. On
the other hand, the limiting value of the QMC data for vanishing quantum
coupling is $0.943$~\cite{RJ96}, displaying a significant disagreement
with the by now established value $0.895$ quoted above for the classical
XY model.

\begin{figure}[b]
\centerline{
\psfig{bbllx=12mm,bblly=15mm,bburx=190mm,bbury=138mm,%
 figure=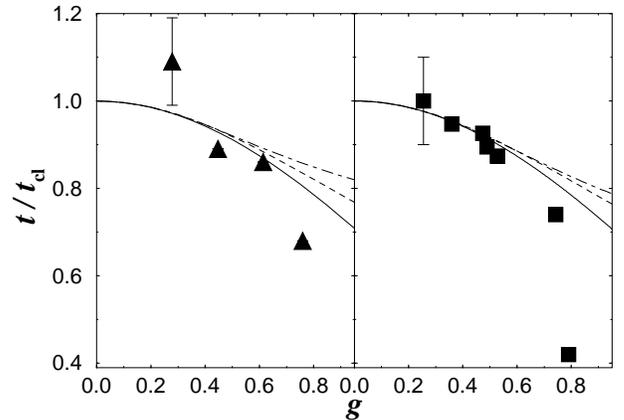,width=80mm,angle=0}} \caption{Phase diagram for the
 quantum XY model, Eq.~(\ref{e.SXY}), on triangular (on the left) and
 square (on the right) lattice in the $ g\!-\!t $ plane at fixed value
 of $\eta=100$. Solid line: $\Gamma_0=\Gamma_1=0$; dashed line:
 $\Gamma_0=3$ and $\Gamma_1=0$; dot-dashed line: $\Gamma_0=0$ and
 $\Gamma_1=3$. The black squares and triangles are the experimental
 data of Ref.~\protect\onlinecite{ZEGM96}.} \label{f.exp}
\end{figure}

In Fig.~\ref{f.exp} the transition temperature versus $g$ is compared with
the experimental data both for square and triangular lattices. It must be
noticed that the {\it a priori} knowledge of the model parameters ($J$,
$C_0$, and $\eta$) describing the experimental setup is rather poor
\cite{ZEGM96}: therefore there is an uncertainty both on the vertical
($t=T/J$) and on the horizontal [$g$ as given by Eqs. (\ref{e.g} and
\ref{e.Omega})] scale. Indeed, the values of $J$ used in
Ref.~\onlinecite{ZEGM96} give extrapolated classical transition
temperatures which are larger than the known theoretical ones, both for
the square ($0.95$ against $0.895$) and the triangular ($1.7$ against
$1.36$) lattice. In order to avoid such systematic error, each set of data
in Fig.~\ref{f.exp} is normalized to the corresponding classical
extrapolated value, as already done in Ref.~\onlinecite{ZEGM96}. The
overall agreement appears to be rather good even though the experiments
present a more rapid decrease of the transition temperature for increasing
values of $g$: this could suggest that the pure XY model is insufficient
to explain this behavior.

Let us now introduce the dissipation. Usually, the damping is described in
terms of shunt resistors connecting the islands to ground or/and shunt
resistors in parallel to the junctions~\cite{jjadiss}. Furthermore the
environmental coupling is taken to be of Ohmic type, i.e.
\begin{equation}
{\hat\gamma}_{\boldsymbol{ij}}(s)=
\gamma_{\boldsymbol{ij}}=\frac{1}{2\pi} R_{\rm Q}
G_{\boldsymbol{ij}}~
\label{e.gammas}
\end{equation}
does not depend on $s$. $R_{\rm Q}=2\pi\hbar/q^2$ is the quantum
resistance and $G_{\boldsymbol{ij}}$ is the conductance
matrix, which reads
\begin{equation}
G_{\boldsymbol{ij}} = \frac{1}{R_0}\delta_{\boldsymbol{ij}}
+\frac{1}{R_1}\,\Big(z\,\delta_{\boldsymbol{ij}}-
\sum_{\boldsymbol{d}}\delta_{\boldsymbol{i},\boldsymbol{j+d}}\Big)~,
\end{equation}
where $R_0$ and $R_1$ are resistive shunts to the ground and
between islands, respectively.

However, a strictly Ohmic damping leads to unphysical results as the
logarithmic ultra-violet divergence of the fluctuations of
momenta~\cite{Weiss99,CFTV99+97}, i.e. the number of Cooper pairs in each
island. The simplest way to consider the inertia in the response of the
dissipation bath is to use the Drude model~\cite{Weiss99}, that consists
of a real-time memory damping with exponential decay of the form
\begin{equation}
 \gamma(t)=\theta(t)~\frac{\gamma}{\tau}~e^{-t/\tau}~,
\label{e.gammaDt}
\end{equation}
where $\theta(t)$ is the step function and $\tau^{-1}$ is the Drude
ultraviolet cutoff frequency; the Ohmic behavior is recovered in the
limit $\tau \ll t$. With this regularization Eq.~(\ref{e.gammas})
becomes
\begin{eqnarray}
{\hat\gamma}_{\boldsymbol{ij}}(s)&=&\frac{R_{\rm Q}}{2\pi}
\bigg[\frac{\delta_{\boldsymbol{ij}}}{R_0}\,
\frac{1}{1+s\,\tau_0}\cr&+&
\frac{1}{R_1}\,\Big( z\,\delta_{\boldsymbol{ij}}-
\sum_{\boldsymbol{d}}\delta_{\boldsymbol{i},\boldsymbol{j+d}}\Big)
\frac{1}{1+s\,\tau_1}\bigg]~.
\label{e.gammasD}
\end{eqnarray}
The presence of two distinct characteristic times is consistent with the
choice of two independent damping mechanism, the on-site and the
nearest-neighbor one, related to $R_0$ and to $R_1$, respectively. The two
characteristic times, $\tau_\ell$ ($\ell=0,1$), have to be compared with
the characteristic times of the equivalent circuit, obtained generalizing
the resistively and capacitively shunted junction model~\cite{Tink96} to a
2D array, i.e. $R_0 C_0$ and $R_1 C_1$. If $\tau_\ell$ is smaller than
$R_\ell C_\ell$ the response of the baths can be considered Ohmic; in the
opposite case the behavior of the system is no longer resistive and the
inertia in the response of the dissipation bath must be considered. In our
calculation, we have therefore assumed $\tau_\ell=R_\ell C_\ell$ as a
representative values of the unknown characteristic times of the baths.

In Fig.~\ref{f.exp} we have also plotted the modification of the phase
diagram due to damping for realistic values of dissipation: we use as
dissipation parameters the dimensionless quantities,
$\Gamma_\ell=R_Q/R_\ell$. The comparison with the experimental data shows
that this kind of dissipation is not very relevant for low values of $g$.
Increasing $g$, but remaining in the range where our approach is valid,
the dissipation appears to affect the phase diagram in the opposite way
with respect to the tendency of the actual experiments. Although our
results improve the quantitative accuracy, these agree with the
qualitative behavior already found in previous works~\cite{jjadiss}.

The problem of a theoretical explanation of the phase diagram of JJA is
thus open, and it might be worthy to investigate whether the common
schematization of a Caldeira-Leggett coupling trough the phase variables,
Eq.~(\ref{e.SD}), is fully justified from a fundamental point of view for
describing a resistive shunt. 
Moreover, it is also known that dissipation does not have a quenching
effect onto the fluctuations of all dynamical variables, as the simple
case of the damped harmonic oscillator shows~\cite{Weiss99}, and it might
be possible to reasonably modify the mechanism of the environmental 
interaction in order to reproduce the observed phase diagram.

We thank J.E.~Mooij and H.S.J.~van~der~Zant for fruitful correspondence,
as well as G. Falci for useful discussions. This work was partly supported
by the project COFIN98 of MURST.




\end{document}